\documentstyle[epsfig,referee]{mn}

\title[]{Optical Flashes and Very Early Afterglows in Wind Environments}

\author[]{X. F. Wu,$^{1 \; \star}$ Z. G. Dai,$^{1 \; \star}$ Y. F. Huang,$^{1,2 \; \star}$
   and T. Lu$^{1,2}$
\thanks{E-mail: xfwu@nju.edu.cn(XFW); daizigao@public1.ptt.js.cn(ZGD); hyf@nju.edu.cn(YFH); tlu@nju.edu.cn(TL)} \\
$^1${\sl Department of Astronomy, Nanjing University, Nanjing
210093,
         P. R. China} \\
$^2${\sl LCRHEA, Institute for High-Energy Physics, Chinese
Academy of
         Sciences, Beijing 100039, P. R. China} }
\date{Accepted ......  Received ......; in original form ......
      }

\begin{document}
\voffset=-0.5 in

\maketitle
\begin{abstract}
The interaction of a relativistic fireball with its ambient medium
is described through two shocks: a reverse shock that propagates
into the fireball, and a forward shock that propagates into the
medium. The observed optical flash of GRB $990123$ has been
considered to be the emission from such a reverse shock. The
observational properties of afterglows suggest that the
progenitors of some GRBs may be massive stars and their
surrounding media may be stellar winds. We here study very early
afterglows from the reverse and forward shocks in winds. An
optical flash mainly arises from the relativistic reverse shock
while a radio flare is produced by the forward shock. The peak
flux densities of optical flashes are larger than 1 Jy for typical
parameters, if we do not take into account some appropriate dust
obscuration along the line of sight. The radio flare always has a
long lasting constant flux, which will not be covered up by
interstellar scintillation. The non-detections of optical flashes
brighter than about 9th magnitude may constrain the GRBs isotropic
energies to be no more than a few $10^{52}$ ergs and wind
intensities to be relatively weak.

\end{abstract}


\section{Introduction}
Gamma-ray bursts (GRBs) have been profoundly understood since the
discovery of afterglows in 1997 (Wijers 1997; Piran 1999; van
Paradijs, Kouveliotou \& Wijers 2000;  M\'esz\'aros 2002). The
hydrodynamic evolution of a GRB remnant is well described by an
ultra-relativistic forward shock that sweeps into an interstellar
medium (ISM) and slows down to the non-relativistic phase while
the resulting afterglow is due to synchrotron radiation of
electrons accelerated by the shock.

However, the emission from a reverse shock propagating into the
shell of the GRB ejecta, was also predicted (M\'esz\'aros $\&$
Rees 1997; Sari $\&$ Piran 1999a). Before the reverse shock
crosses the shell, the shocked shell matter carries an amount of
energy comparable to that of the forward shocked interstellar
medium. The prompt optical flash of GRB $990123$ (Akerlof et al.
1999) motivates investigations of emission from the reverse shock
(M\'esz\'aros $\&$ Rees 1999; Sari $\&$ Piran 1999b; Kobayashi
$\&$ Sari 2000; Kobayashi 2000). The light curve of an optical
flash in a uniform ISM has been well studied in both thin shell
and thick shell cases (Kobayashi 2000) and it is found that the
optical flash of GRB $990123$ is due to the radiation from a
mildly relativistic reverse shock (Kobayashi $\&$ Sari 2000).
Recently, the contribution of reverse shocks to early radio
afterglows was also discussed by Soderberg $\&$ Ramirez-Ruiz
(2002).

There is increasing evidence for the association between some GRBs
and type Ib/c supernovae. The most direct evidence is the
association of GRB $980425$ with SN $1998$bw (Galama et al. 1998).
This association is strengthed by its radio emission after the
burst (Li $\&$ Chevalier 1999). Up to date several GRB afterglows,
e.g., GRBs $970228$ (Reichart 1999), $970508$ (Sokolov 2001),
$980326$ (Bloom et al 1999), $990712$ (Sahu et al. 2000), $991208$
(Castro-Tirado et al. 2001), $000911$ (Lazzati et al. 2001), and
$011211$ (Bloom et al. 2002), show a supernova component in their
light curves peaking at $t_{\oplus}=(1+z)t_{\rm SN,peak}$, with
$t_{\rm SN,peak}\sim$ 15 days for SN Ic. One of the most popular
central engines of GRBs now is believed to be the collapse of
massive stars (Woosley 1993; Paczy\'nski 1998). This
collapsar/hypernova scenario naturally explains the broad and
shifted iron-group emission lines in GRBs (McLaughlin, Wijers $\&$
Brown 2002). The iron line emission can be attributed in the
envelope models either from the irradiated shallower layers on the
funnel wall along the rotation axis by a continuing but decaying
relativistic outflow (Rees $\&$ M\'esz\'aros 2000) or from the
dissipation of the magnetized thermal bubble expanding into H
envelope of the progenitor, which results from the interaction
between the jet with continuing energy injection and the stellar
envelope (M\'esz\'aros $\&$ Rees 2001). The place of the iron line
emission in the envelope models is at H envelope radius $r\sim
10^{13}$ cm, where a small amount of iron, $\sim 10^{-8}$
M$_{\odot}$ and $\sim 10^{-5}$ M$_{\odot}$, can meet the
requirement. Other iron line emission models require a large line
emission radius, $r\sim 10^{16}$ cm, and much larger iron mass
$\geq 0.06$ M$_{\odot}$ (Vietri et al. 2001; B\"{o}ttcher $\&$
Fryer 2001). The measured emission lines in the X-ray afterglow of
GRB 011211 favors the latter (Reeves et al. 2002). The
pre-explosion wind structure can survive in envelope models but
may be destroyed in the other models due to either the ejection of
SN prior to GRB or the common envelope evolution in the binary
system. Detailed calculations and more sensitive data on X-ray
afterglow spectral features will distinguish between the various
line emission models (Ballantyne $\&$ Ramirez-Ruiz 2001).

The light curve steepening of some well-observed GRB afterglows
deviates from the predictions of the standard model (Sari, Piran
$\&$ Narayan 1998). This deviation leads to discussions of the
environment and geometry effects. Previous studies have discussed
the afterglow from a spherical blast wave in a wind (Dai $\&$ Lu
1998; Chevalier $\&$ Li 1999, 2000) and from a jet in a wind
(Livio $\&$ Waxman 2000; Dai $\&$ Gou 2001; Gou et al. 2001). The
wind environments of GRBs $970228$, $970508$, $991216$ and
$000301c$ are recognized in the spherical-wind model (Chevalier
$\&$ Li 2000; Li $\&$ Chevalier 2001). In the jet model, ten
broadband GRB afterglows (i.e., GRBs 970508, 980519, 990123,
990510, 991208, 991216, 000301c, 000418, 000926, and 010222) have
been well fitted (Panaitescu 2001; Panaitescu $\&$ Kumar 2001,
2002). Among them, only GRBs $980519$, $990123$ and $990510$ are
inconsistent with wind environments. However, we also noticed that
Frail et al. (2000) gave a better fit for their first 63 days
observations of the radio afterglow of GRB $980519$ in the
spherical-wind model.

Information on the surrounding environment of a GRB can also be
obtained by considering the evolution of the soft X-ray absorption
property according to the surrounding density distribution
(Lazzati $\&$ Perna 2002). The expected absorption of a wind
environment will quickly drop to the undetectable level in no more
than 1 second. It conflicts with the evidence of variable soft
X-ray absorption in three GRBs detected by \textit{Beppo}SAX.
Nonetheless, the wind environments can not be excluded by this
method due to the limited samples. It may give some evidence for
another type of GRBs environments, such as dense clouds.
Superbubbles which are proposed as the sites of GRBs have been
expected to explain the obtained low densities $n\sim 10^{-3}$
cm$^{-3}$ (Scalo $\&$ Wheeler 2001). It requires a very weak
stellar wind at least 6 orders weaker than that of the progenitor
of SN $1998$bw to insure non-contamination of the supperbubble by
the wind at a typical afterglow radius. The wind profile $n\propto
r^{-2}$ may be terminated at some radius $r\sim 10^{17}$ cm, where
the wind sweeps up a mass of ISM comparable to that of the wind
and leads to a quasi-homogeneous dense shell with the mean density
$\sim$ $10^2$ cm$^{-3}$ - $10^3$ cm$^{-3}$ and the thickness
$\sim$ $10^{17}$ cm (Ramirez-Ruiz et al. 2001). This termination
of the wind profile will not affect our analysis of the dynamic
evolution of very early afterglows, but has the potential
reduction of the optical flash. In this paper we perform a more
detailed calculation of the very early afterglows of GRBs in wind
environments including the reverse shock effect. In \S 2 we give
the hydrodynamics of forward and reverse shocks. In \S 3 early
afterglow light curves at optical and radio bands are calculated.
We discuss some observational implications of our results in \S 4.
The synchrotron self-absorption of electrons with a broken power
law energy distribution is discussed in the appendix.

\section{Hydrodynamics of forward and reverse shocks}

We consider a uniform and cold relativistic coasting shell with
rest mass $M_0$, the energy $E$, the initial Lorentz factor
$\eta=E/M_0 c^2$, and the observed width $\Delta$, which sweeps up
a wind with the particle number density $n_1=Ar^{-k}$ ($k=2$),
where $A=\dot{M}/4 \pi m_p v_w=3\times 10^{35}A_{\ast} {\rm
cm}^{-1}$ and $A_{\ast}=(\dot{M}/1\times
10^{-5}M_{\odot}yr^{-1})(v_w/10^3 {\rm km/s})^{-1}$ (Chevalier
$\&$ Li 1999, 2000). Two shocks develop: a forward shock
propagating into the wind and a reverse shock propagating into the
shell. There are four regions separated by the two shocks: (1) the
unshocked wind, (2) the shocked wind, (3) the shocked shell
material, (4) the unshocked shell material. From the shock jump
conditions and equality of pressures and velocities along the
contact discontinuity, the Lorentz factor $\gamma$, the pressure
$p$, and the number density $n$ in both shocked regions can be
determined by $n_1$, $n_4$, and $\eta$ (Blandford $\&$ McKee 1976,
hereafter BM).

We define  the Sedov radius, at which the swept-up wind's
rest-mass energy equals the initial energy $E$ of the shell, i.e.,
$l=M_0\eta/(4\pi A m_p)$. Assuming that the shell doesn't enter
the spreading stage, the width $\Delta$ is a constant, and
$n_4=M_0/(4\pi m_p r^2 \Delta \eta) \propto r^{-2}$. The property
of the reverse shock is largely determined by a parameter defined
as $f=n_4/n_1$ (Sari $\&$ Piran 1995), which can be further given
by
\begin{equation}
f=\frac{l}{\Delta \eta^2}.
\end{equation}

As shown by Sari $\&$ Piran (1995), if $f \gg \eta^2$, or the
baryon loading $M_0 \gg 4\pi m_p A \Delta \eta^3$, the reverse
shock will be a Newtonian reverse shock (NRS), and the reverse
shock will be a relativistic reverse shock (RRS) if $f \ll \eta^2$
or $M_0 \ll 4\pi m_p A \Delta \eta^3$. For RRS, the Lorentz factor
of the shocked shell relative to the unshocked shell is
$\overline{\gamma}_3=\eta^{1/2} f^{-1/4}/\sqrt{2}$, and the
Lorentz factors of shocked matter are
$\gamma_2=\gamma_3=\eta^{1/2} f^{1/4}/\sqrt{2}$. The time for  the
reverse shock to cross the shell in the fixed frame is
\begin{equation}
t_{\Delta}={\alpha \Delta \eta f^{1/2}\over c},
\end{equation}
where $\alpha=1/2$ for RRS and $\alpha=3/\sqrt{14}$ for NRS (Sari
$\&$ Piran 1995). The shock expands to the radius
$R_{\Delta}\approx c t_{\Delta}=\alpha \sqrt{l \Delta}$ when the
reverse shock is just over. In addition to $R_{\Delta}$ there are
two other characteristic radii involved here: the deceleration
radius $R_{\eta}=l \eta^{-2}$, at which the swept-up mass is
$M_0/\eta$, and the shell spreading radius $R_s=\Delta \eta^2$.
They have different dependences on $l$, $\Delta$ and $\eta$ for
the wind and ISM cases (Sari $\&$ Piran 1995). A parameter
$\zeta=\eta^{-2}\sqrt{l/\Delta} $ is introduced to compare these
three radii. According to equation (1), we get
\begin{equation}
R_{\eta}=\zeta R_{\Delta}/\alpha=\zeta^2 R_s,
\end{equation}
with $f=\zeta^2\eta^2$. $\zeta\ll1$ describes RRS: $f\ll \eta^2$,
and $R_{\eta}\ll R_{\Delta} \ll R_s$, while $\zeta\gg1$ describes
NRS: $f\gg \eta^2$, and $R_s\ll R_{\Delta} \ll R_{\eta}$. For RRS,
the previous assumption of $\Delta=$const is always satisfied, and
deceleration of the shell happens before the reverse shock crosses
it. For NRS we must consider the spreading effect of the shell,
and $\Delta=r/\eta^2$ with $n_4=A l/r^3$. The comoving density
ratio $f=l/r$ decreases with increasing $r$. The reverse shock becomes
trans-relativistic at $R_N=l/\eta^2$(=$R_{\eta}$) when $f=\eta^2$.
A similar correction on $t_{\Delta}$ and $R_{\Delta}$ must be done
since $\Delta$ and $f$ vary with $r$. We are most interested in
RRS because of both a large energy output and a constant width.

Before RRS crosses the shell, the distance that the shell travels
when RRS crosses a length $dx$ in the unshocked shell frame can be
given by (Kobayashi 2000),
\begin{equation}
dr=\alpha \eta \sqrt{f} dx.
\end{equation}
The shocked electron number is proportional to $x/\Delta$. In the
observer's frame we have $r=2c\gamma_3^2t_{\oplus}$, so the
hydrodynamic variables are
\begin{equation}
n_3=\frac{8\sqrt{2}A}{\eta
l^{1/4}\Delta^{7/4}}\left(\frac{t_{\oplus}}{T}\right)^{-2}, \,\,\,\,
e_3=\frac{8Am_p c^2}{l^{1/2} \Delta^{3/2}}\left(\frac{t_{\oplus}}{T}\right)^{-2},
\end{equation}
\begin{equation}
\gamma_3=\frac{1}{\sqrt{2}}\left(\frac{l}{\Delta}\right)^{1/4},\,\,\,\,
N_e=N_0\frac{t_{\oplus}}{T},
\end{equation}
where $N_0=M_0/m_p$ is the total electron number of the shell and
$T=\Delta/(2c)$ is the observer's time at which the RRS crosses the
shell, which is one half of the crossing time adopted by Kobayashi
(2000).

After RRS crosses the shell, the temperature of the shocked shell is
very high and it will expands adiabatically and enters
the spreading stage. Since it locates not far away from the forward
shocked wind, we take the BM self-similar adiabatic
solution with impulsive energy injection for the
evolution of the shocked shell (Kabayashi $\&$ Sari 2000). With
$p_3\propto n_3^{4/3}$, we get $p_3\propto r^{(4k-26)/3}\propto
r^{-6}, \gamma_3\propto r^{(2k-7)/2}\propto r^{-3/2}, n_3\propto
r^{(2k-13)/2}\propto r^{-9/2}$ and $r\propto t_{\oplus}^{1/4}$.
The hydrodynamic variables are therefore given by
\begin{equation}
n_3=n_3(T)\left(\frac{t_{\oplus}}{T}\right)^{-9/8},\,\,\,\,
e_3=e_3(T)\left(\frac{t_{\oplus}}{T}\right)^{-3/2},
\end{equation}
\begin{equation}
\gamma_3=\gamma_3(T)\left(\frac{t_{\oplus}}{T}\right)^{-3/8},\,\,\,\,
N_e=N_0.
\end{equation}

The hydrodynamic evolution of a relativistic forward shock in a
wind has already been discussed in some details (Dai $\&$ Lu 1998;
Chevalier $\&$ Li 2000; Panaitescu $\&$ Kumar 2000). Before RRS
crosses the shell, the Lorentz factor of forward shock
$\gamma_{sh}=\sqrt{2}\gamma_2$ is a constant and equals
$\sqrt{2}\gamma_3$. The forward shock radius is $r=2\gamma_{sh}^2
c t_{\oplus}=4\gamma_2^2 c t_{\oplus}$. Note that there is a
difference of a factor of 2 between the reverse shock radius and
the forward shock radius, which arises from the fact that the
forward shock moves with $\gamma_{sh}$ at any time although the
photon emitter travels with $\gamma_2$ (Chevalier $\&$ Li 2000).
So  the time when the reverse shock is finished becomes
$T^{\prime}=T/2$. For $t_{\oplus}\leq T^{\prime}$
\begin{equation}
n_2=\frac{8\sqrt{2}A}{l^{3/4}\Delta^{5/4}}\left(\frac{t_{\oplus}}{T^{\prime}}
\right)^{-2},\,\,\,\, e_2=\frac{8A m_p c^2}{l^{1/2}
\Delta^{3/2}}\left(\frac{t_{\oplus}}{T^{\prime}}\right)^{-2},
\end{equation}
with $\gamma_2=\sqrt{2}(l/\Delta)^{1/4}/2$. For $t_{\oplus} >
T^{\prime}$, the hydrodynamic variables evolve as
\begin{equation}
n_2=n_2(T^{\prime})\left(\frac{t_{\oplus}}{T^{\prime}}\right)^{-5/4},\,\,\,\,
e_2=e_2(T^{\prime})\left(\frac{t_{\oplus}}{T^{\prime}}\right)^{-3/2},\,\,\,\,
\gamma_2=\gamma_2(T^{\prime})\left(\frac{t_{\oplus}}{T^{\prime}}\right)^{-1/4}.
\end{equation}
The above analytical hydrodynamic evolution for both the reverse
shock and forward shock is valid only in the relativistic phase.
However, a late-time afterglow may be produced by a
non-relativistic forward shock (Huang, Dai $\&$ Lu 1998, 1999,
2000). Thus, our results are applied only for early or very early
afterglows in wind environments, including the reverse shock
emission.

\section{Light curve of very early afterglow}
\subsection{Radiation from RRS}
The shock accelerates electrons to a power-law energy distribution
with an index $p$ and with a minimum Lorentz factor
$\gamma_m=\xi_e (\overline{\gamma}_3-1)[(p-2)/(p-1)] m_p/m_e$,
where $\xi_e$ is a fraction of the internal energy density carried
by the electrons. The standard radiation mechanism for GRB
afterglows is synchrotron radiation by relativistic electrons in
the magnetic field whose energy density is another fraction
($\xi_B$) of the internal energy density. We also consider
synchrotron self-absorbtion at the radio band. Thus, the spectrum
as discussed by Sari, Piran $\&$ Narayan (1998) in the ISM case
can be used to discuss an afterglow in the wind case. The spectrum
is characterized by three break frequencies: the typical frequency
$\nu_m$, the cooling frequency $\nu_c$, and the self-absorption
frequency $\nu_a$. A detailed derivation of $\nu_a$ for
fast-cooling and slow-cooling regimes is given in the appendix.
Since $B\propto (\xi_B n_1)^{1/2} \gamma_3$  and $\gamma_c\propto
1/(\gamma_3 B^2 t_{\oplus})$, we have $\nu_m\propto \gamma_3
\gamma_m^2 B$ and $\nu_c\propto \gamma_3 \gamma_c^2 B$. The peak
flux $F_{\nu,max}\propto N_e \gamma_3 B/D^2$, where $D$ is the
distance from the GRB site to the observer.

Before RRS crosses the shell ($t_{\oplus}\leq T$),
\begin{equation}
\nu_m=8.65\times
10^{16}E_{52}^{-1/2}\xi_{e,0}^2\xi_{B,-2}^{1/2}\eta_{300}^2
A_{\ast}\Delta_{13}^{-1/2}\left(\frac{t_{\oplus}}{T}\right)^{-1}\,{\rm Hz},
\end{equation}
\begin{equation}
\nu_c=1.32\times 10^{12}E_{52}^{1/2}\xi_{B,-2}^{-3/2}
A_{\ast}^{-2}\Delta_{13}^{1/2}\frac{t_{\oplus}}{T}\,{\rm Hz},
\end{equation}
\begin{equation}
F_{\nu,max}^{rs}=1.42E_{52}\xi_{B,-2}^{1/2}\eta_{300}^{-1}
A_{\ast}^{1/2}\Delta_{13}^{-1} D_{28}^{-2}\,{\rm Jy},
\end{equation}
where $E=10^{52}E_{52}$ ergs, $\eta=300\eta_{300}$,
$\Delta=10^{13}\Delta_{13}$ cm, $\xi_B=10^{-2}\xi_{B,-2}$,
$\xi_e=0.6\xi_{e,0}$, $A=3\times10^{35}A_{\ast}\,{\rm cm}^{-1}$
and $D=10^{28}D_{28}$ cm. The time for the reverse shock to cross
the shell is $T=167\Delta_{13}$ s. For $\nu_c\leq\nu_a\leq\nu_m$,
we have
\begin{equation}
\nu_a=4.11\times 10^{13}E_{52}^{1/6}\eta_{300}^{-1/3}
A_{\ast}^{1/6}\Delta_{13}^{-5/6}\left(\frac{t_{\oplus}}{T}\right)^{-2/3}\,
{\rm Hz},
\end{equation}
which is consistent with eq. (18) of Dai $\&$ Lu (2001). For
$\nu_a\leq\nu_c\leq\nu_m$, we have
\begin{equation}
\nu_a=6.47\times 10^{14}E_{52}^{-1/10}\eta_{300}^{-3/5}A_{\ast}^{19/10}
\Delta_{13}^{-19/10}\xi_{B,-2}^{6/5}\left(\frac{t_{\oplus}}{T}\right)^{-2}\,{\rm Hz}.
\end{equation}
The cooling Lorentz factor ($\gamma_c\propto t_{\oplus}/T$) is
determined by the dynamical timescale. At very early times,
$\gamma_c$ may be less than unity, which is impossible in physics.
In fact the radiation power of such electrons with $\gamma_c\simeq
1$ will be cyclotron power, instead of synchrotron power.
According to Dai \& Lu (2001), we neglect the radiation before
$t_{\oplus,{\rm crit}}$ (same as $t_0$ in Dai $\&$ Lu 2001) when
$\gamma_c=1$,
\begin{equation}
t_{\oplus,{\rm
crit}}=11.3E_{52}^{-1/4}\Delta_{13}^{1/4}\xi_{B,-2}A_{\ast}^{5/4}\,{\rm
s}
\end{equation}
which is the beginning for our calculation.

When RRS crosses the shell ($t_{\oplus}> T$), the
shock-accelerated electrons will be in the slow cooling regime if
$\nu_c(T)>\nu_m(T)$. However, there will be no radiation if
$\nu_c(T)<\nu_m(T)$, because all the previously shocked electrons
have cooled and no newly-shocked electrons are in the shell when
$t_{\oplus}>T$. The break frequencies and peak flux for
$\nu_c(T)>\nu_m(T)$ are
\begin{equation}
\nu_m=\nu_m(T)\left(\frac{t_{\oplus}}{T}\right)^{-15/8},
\nu_{cut}=\nu_c(T)\left(\frac{t_{\oplus}}{T}\right)^{-15/8},
F_{\nu,max}=F_{\nu,max}(T)\left(\frac{t_{\oplus}}{T}\right)^{-9/8}.
\end{equation}
We have also derived the self-absorption frequency:  in the slow cooling phase,
$\nu_a\propto(t_{\oplus}/T)^{-3/5}$  for $\nu_a\leq\nu_m\leq\nu_{cut}$, and
 $\nu_a\propto(t_{\oplus}/T)^{-(15p+26)/8(p+4)}$ for
$\nu_m\leq\nu_a\leq\nu_{cut}$. The time for  the reverse shocked
shell to become non-relativistic, i.e., when $\gamma_3=2$, is
\begin{equation}
t_{\oplus,nr}=197.0 E_{52}^{2/3} A_{\ast}^{-2/3}\Delta_{13}^{-2/3} T.
\end{equation}

\subsection{Radiation from the forward shock}
Before RRS crosses the shell, the forward shocked wind matter
moves with the same Lorentz factor as the reverse shocked shell
matter. Whilst most of the initial kinetic energy of the shell has
been transferred to the forward shocked matter, as shown in eq.
(3). The forward shock decelerates at a slower rate than the
shocked shell does after the crossing time. We can consider the
shocked shell as a tail of the forward shock based on the BM
solution with its self-similar parameter $\chi\propto
r^{4-k}\propto r^2$, which describes inward motion of the shocked
shell relative to the shock front as the shock radius increases.

Before RRS crosses the shell ($t_{\oplus}\leq T^{\prime}$), the
radiation from the forward shock can be described by
\begin{equation}
\nu_m=1.70\times
10^{17}E_{52}^{1/2}\xi_{e,0}^2\xi_{B,-2}^{1/2}\Delta_{13}^{-3/2}
\left(\frac{t_{\oplus}}{T^{\prime}}\right)^{-1}\,{\rm Hz},
\end{equation}
\begin{equation}
\nu_c=5.28\times 10^{12}E_{52}^{1/2}\xi_{B,-2}^{-3/2}
A_{\ast}^{-2}\Delta_{13}^{1/2}\frac{t_{\oplus}}{T^{\prime}}\,{\rm Hz},
\end{equation}
\begin{equation}
F_{\nu,max}^{fs}=5.07E_{52}^{1/2}A_{\ast}\Delta_{13}^{-1/2}
D_{28}^{-2}\,{\rm Jy}.
\end{equation}
For $\nu_c\leq\nu_a\leq\nu_m$, we have the self-absorption frequency
\begin{equation}
\nu_a=2.92\times
10^{13}A_{\ast}^{1/3}\Delta_{13}^{-2/3}\left(\frac{t_{\oplus}}{T^{\prime}}\right)^{-2/3}\,{\rm Hz},
\end{equation}
and for $\nu_a\leq\nu_c\leq\nu_m$ we have
\begin{equation}
\nu_a=9.78\times
10^{13}E_{52}^{-2/5}A_{\ast}^{11/5}\Delta_{13}^{-8/5}\xi_{B,-2}^{6/5}
\left(\frac{t_{\oplus}}{T^{\prime}}\right)^{-2}\,{\rm Hz}.
\end{equation}
The time for $\gamma_c\geq 1$ is
\begin{equation}
t_{\oplus,crit}^{\prime}=2.81E_{52}^{-1/4}\Delta_{13}^{1/4}\xi_{B,-2}A_{\ast}^{5/4}\,{\rm s},
\end{equation}
which is the beginning of our calculation. Since the peak flux of
the forward shock emission is comparable to that of the RRS
emission, a very early afterglow may be dominated by both of the
two regions. An optical flash may even be dominated by the RRS
emission if $E$ is very large and $A_{\ast}$ is small, which can
be estimated by comparing eq. (13) with eq. (21).

The break frequencies and peak flux for $t_{\oplus}>T^{\prime}$
are
\begin{equation}
\nu_m=\nu_m(T^{\prime})\left(\frac{t_{\oplus}}{T^{\prime}}\right)^{-3/2},
\nu_c=\nu_c(T^{\prime})\left(\frac{t_{\oplus}}{T^{\prime}}\right)^{1/2},
F_{\nu,max}=F_{\nu,max}(T^{\prime})\left(\frac{t_{\oplus}}{T^{\prime}}\right)^{-1/2}.
\end{equation}
The self-absorption frequency is: (1) in the fast cooling phase,
$\nu_a\propto(t_{\oplus}/T^{\prime})^{-2/3}$ for
$\nu_c\leq\nu_a\leq\nu_m$, and
$\nu_a\propto(t_{\oplus}/T^{\prime})^{-8/5}$ for
$\nu_a\leq\nu_c\leq\nu_m$; (2) in the slow cooling phase,
$\nu_a\propto(t_{\oplus}/T^{\prime})^{-3/5}$ for
$\nu_a\leq\nu_m\leq\nu_c$.  The case of $\nu_m\leq\nu_a\leq\nu_c$
seems impossible before $t_{\oplus,nr}^{\prime}=2.763\times10^3
E_{52}A_{\ast}^{-1} \Delta_{13}^{-1} T^{\prime}$.

\subsection{Light curves of optical and radio afterglows}
The above subsections give the spectral profiles and the decaying
laws of the break frequencies. In this subsection, we will discuss
the light curves at some observed frequencies such as $\nu_{opt}=
4\times10^{14}$ Hz and $\nu_{rad}=8.46$ GHz. We calculate the
evolution of the flux densities at these two frequencies with
fixed parameters of $\xi_{e,0}=1.0$, $\xi_{B,-2}=1.0$ and
$\Delta_{13}=0.5$ but with $E_{52}$ varying between 1.0 and
$10^2$, and $A_{\ast}$ varying between $10^{-2}$ and 1, to
investigate the effects of energy and wind intensity. The values
of $\xi_{e,0}$ and $\xi_{B,-2}$ are consistent with the afterglow
of GRB $970508$ (Wijers $\&$ Galama 1999; Granot, Piran $\&$ Sari
1999) and the optical flash and afterglow of GRB $990123$ (Wang,
Dai $\&$ Lu 2000), and $\Delta_{13}=0.5$ is consistent with
$T^{\prime}=\Delta/4c\approx 42$ sec, which is the peak time of
the optical flash of GRB $990123$. The light curves in six cases
and at two frequencies are shown in Figures 1-6.

The emission from RRS is nearly comparable or dominant over that
from the forward shock at optical frequency. This is naturally
estimated from the ratio of the reversely shocked electron number
of the shell to the forward shocked electron number,
\begin{equation}
\frac{N_{e,rs}}{N_{e,fs}}=3f^{1/2}=4.21E_{52}^{1/2}A_{\ast}^{-1/2}\Delta_{13}^{-1/2}\eta_{300}^{-1},
\end{equation}
before RRS crosses the shell, compared with eq.(58) of Chevalier
$\&$ Li (2000). In detail, the maximum optical flux in RRS reaches
at $t_{\oplus}=T$,
\begin{eqnarray}
F_{\nu_{opt}}^{rs}(T)&=&F_{\nu,max}^{rs}(\frac{\nu_{opt}}{\nu_c})^{-1/2}\nonumber\\
                     &=&8.16\times 10^{-2}
                     E_{52}^{5/4}\eta_{300}^{-1}A_{\ast}^{-1/2}\Delta_{13}^{-3/4}\xi_{B,-2}^{-1/4}D_{28}^{-2}{\rm Jy},
\end{eqnarray}
when RRS is still in the fast cooling regime, as emphasized by
Chevalier $\&$ Li (2000). The above equation is not suitable for
the case of $E_{52}=10, A_{\ast}=10^{-2}$, where RRS enters the
slow cooling regime before it crosses the shell, because we can
use eqs.(11) and (12) to get the transition time $t_{mc}^{rs}$ by
$\nu_m=\nu_c$,
\begin{equation}
\frac{t_{mc}^{rs}}{T}=2.56\times 10^2
E_{52}^{-1/2}\eta_{300}A_{\ast}^{3/2}\Delta_{13}^{-1/2}\xi_{e,0}\xi_{B,-2}=0.11<1,
\end{equation}
for $E_{52}=10, A_{\ast}=10^{-2}$. The optical flux contributed by
the forward shock reaches its maximum at $t_{\oplus}=T^{\prime}$,
\begin{eqnarray}
F_{\nu_{opt}}^{fs}(T^{\prime})&=&F_{\nu,max}^{fs}(\frac{\nu_{opt}}{\nu_c})^{-1/2}\nonumber\\
                     &=&0.583E_{52}^{3/4}\Delta_{13}^{-1/4}\xi_{B,-2}^{-3/4}D_{28}^{-2}{\rm Jy},
\end{eqnarray}
except in the two cases of $E_{52}=10, A_{\ast}=0.1$ and
$E_{52}=10, A_{\ast}=10^{-2}$. For these two cases, the optical
flux by the forward shock reaches the maximum before RRS has
crossed the shell and decays to
\begin{eqnarray}
F_{\nu_{opt}}^{fs}(T^{\prime})&=&F_{\nu,max}^{fs}(\frac{\nu_{opt}}{\nu_c})^{1/3}\nonumber\\
                     &=&1.583E_{53}^{1/3}(\frac{\Delta_{13}}{0.5})^{-2/3}
                     (\frac{A_{\ast}}{0.1})^{5/3}\xi_{B,-2}^{1/2}D_{28}^{-2}{\rm
                     Jy}.
\end{eqnarray}
For more energetic fireball and weaker stellar wind, the emission
from RRS becomes more stronger than that from the forward shock.
The weakest RRS case is for $E=10^{52}$ ergs, $A_{\ast}=1.0$,
where the forward shock dominates RRS at $\nu_{opt}$ by about one
order of magnitude (Fig. 2.). The peak flux $F_{\nu_{opt},{peak}}$
increases with both $E$ and $A_{\ast}$ and varies from $\sim 1$ Jy
to $\sim 10^2$ Jy. An optical flash occurs in all cases, although
its properties such as the brightening and decaying timescales and
the light curve index may be different from each other. The time
scale of emission with flux above 1 Jy ranges from 10  to $10^4$
seconds. None of the six cases discussed here is consistent with
the rapid brightening ($F_{\nu_{opt}}\propto t_{\oplus}^{3.4}$)
and fast decaying ($\propto t_{\oplus}^{-2.1}$) of the optical
flash of GRB $990123$. This further confirms the result of
Chevalier $\&$ Li (2000) that the optical flash of GRB $990123$
originates from a homogeneous ISM.

The radio afterglow is mainly attributed to radiation from the
forward shock and we can neglect the contribution of the emission
from RRS.  The light curve indices of the forward shock are $2$,
$-2/3$, $0$, $-(3p-1)/4$ from low- to high-frequencies in Figs. 1,
4 and 6, as shown in Chevalier $\&$ Li (2000). We also get an
index $\beta=3$, which does not equal to $7/4$ of Chevalier $\&$
Li (2000) at very early times. Indices of 2, 1, and 0 in Fig. 2.
and Fig. 5. are consistent with the results of Chevalier $\&$ Li
(2000). The peak flux at radio band varies by two orders of
magnitude, from $10^3$ to $10^5$ $\mu$Jy. There is a long lasting
platform in the radio flux before the forward shock enters the
non-relativistic phase ( also see Fig. 3 of Panaitescu $\&$ Kumar
2000). It might be used to distinguish the wind environment from
the uniform ISM environment of GRBs and give further hints on the
central engine.

\section{Summary}
We have analyzed very early afterglows in wind environments by
considering the radiation from both a relativistic reverse shock
and a relativistic forward shock. The resulting optical flash in
the wind case is mainly attributed to RRS because of reduction of
the flux from the forward shock by synchrotron self-absorption. Fo
more energetic fireball and weaker stellar wind, the optical
emission from RRS becomes more stronger than that from the forward
shock. The resulting radio flare is largely produced by the
forward shock emission.

Theoretically, the peak flux of the optical flash ranges from 1 Jy
to $10^2$ Jy, and the peak time since the GRB trigger is tens of
seconds, which is determined by the shell width. In our cases we
choose the width according to the typical time scales of long
gamma-ray bursts, i.e., $\sim$ a few tens of seconds. Such a large
optical flux at very early times is due to the dense wind at small
radius. If optical observations could be done in a few seconds
after the GRB trigger, then one can discriminate the wind
environment from the uniform ISM environment. It is a challenge
for current observational instruments. While strong optical
flashes have been predicted theoretically in this research as well
as by many other authors, in realistic observations optical flash
has been observed only from GRB $990123$ till now. The missing of
a significant number of strong optical flashes has been
interpreted as due to dust obscuration (Soderberg $\&$
Ramirez-Ruiz 2002). We propose that further contribution to the
reduction of the optical flash may come from the dense shell
around the stellar wind with the column density about
$10^{20}-10^{21}$ cm$^{-2}$ (Ramirez-Ruiz et al. 2001). Another
characteristic of wind environment which might be easier for
observations is a long lasting platform in radio flux, which
arises from the interaction of the forward shock with the wind
matter.

However, dust extinction could not darken a 4th magnitude ($\sim$
100 Jy) optical flash significantly to be immune to the optical
survey. The non-detections of optical flashes brighter than $\sim$
9th magnitude would eventually constrain the isotropic energies of
GRBs to be no more than a few $10^{52}$ ergs and result in
relatively weak circum-burst wind intensities with $A_{\ast}<1.0$,
according to eqs. (13), (21) and (27). We expect future
observations on early afterglows to diagnose the environments of
GRBs efficiently.

\section*{Acknowledgments}
We wish to thank the referee for valuable suggestions. XFW would
like to  thank D. M. Wei, X. Y. Wang, Z. Li, Y. Z. Fan, and J. B.
Feng for their fruitful discussions. This work was supported by
the National Natural Science Foundation of China (grant number
19973003, 10233010, 10003001), the National 973 project (NKBRSF
G19990754), the Foundation for the Author of National Excellent
Doctorial Dissertation of P. R. China (Project No: 200125) and the
Special Funds for Major State Basic Research Projects.

\begin{appendix}
\section{Synchrotron self-absorption by electrons with a broken
power-law distribution}
\subsection{Synchrotron self-absorption coefficient by electrons
with one single power-law distribution} All the quantities below
are in the comoving reference frame. The power law distribution of
the shock-accelerated electrons can be written as
$N(\gamma_e)d\gamma_e=N_{\gamma_e}\gamma_e^{-p}d\gamma_e$, where
$N_{\gamma_e}$ is the normalized coefficient and
$\gamma_1\leq\gamma_e\leq\gamma_2$. The power radiated from an
electron with $\gamma_e$ is (Rybicki $\&$ Lightman 1979)
\begin{equation}
P(\nu, \gamma_e)=\frac{2\pi\sqrt{3}q_e^2\nu_L
\sin{\theta}}{c}\frac{\nu}{\nu_c}\int_{\nu/\nu_c}^{\infty}K_{5/3}(t)dt,
\end{equation}
where $\nu_L=q_e B/2\pi m_e c$ is the Lamor frequency and
$\nu_c=(3\gamma_e^2\nu_L\sin{\theta})/2$ is the typical frequency
emitted by this electron. We rewrite the self-absorption
coefficient of eq. (6.52) of Rybicki $\&$ Lightman (1979), and get
\begin{equation}
k_{\nu}=-\frac{1}{8\pi m_e \nu^2}\int_{\gamma_1}^{\gamma_2}P(\nu,
\gamma_e) \gamma_e^2
\frac{d}{d\gamma_e}\left(\frac{N(\gamma_e)}{\gamma_e^2}\right)d\gamma_e.
\end{equation}
Before integrating the above equation, we define
$\nu_1=(3\gamma_1^2 \nu_L \sin{\theta})/2$ and $\nu_2=(3\gamma_2^2
\nu_L \sin{\theta})/2$. For $\nu\ll\nu_1$, we have
\begin{equation}
k_{\nu}=\frac{\pi^{1/3}}{2^{2/3}3^{1/3}\Gamma(\frac{1}{3})}\frac{p+2}
{p+\frac{2}{3}}\frac{q_e^{8/3}}{(m_ec)^{5/3}}N_{\gamma_e}
B^{2/3}(\sin{\theta})^{2/3}\gamma_1^{-(p+2/3)}\nu^{-5/3}.
\end{equation}
For simplicity, we take an isotropic distribution of $\theta$ with
$<(\sin{\theta})^{2/3}>=\frac{\sqrt{\pi}\Gamma(\frac{1}{3})}{5\Gamma(\frac{5}{6})}$
and obtain
\begin{equation}
k_{\nu}=114.65\times\frac{p+2}{p+\frac{2}{3}}N_{\gamma_e}B^{2/3}\gamma_1^{-(p+2/3)}\nu^{-5/3}.
\end{equation}
For $\nu_1\ll\nu\ll\nu_2$, we obtain
\begin{equation}
k_{\nu}=g(p)\frac{q_e^{3}}{2\pi m_e }\left(\frac{3q_e}{2\pi m_e^3
c^5 }\right)^{p/2}(m_e c^2)^{p-1}
N_{\gamma_e}B^{(p+2)/2}(\sin{\theta})^{(p+2)/2}\nu^{-(p+4)/2},
\end{equation}
where
$g(p)=\frac{\sqrt{3}}{16}\Gamma(\frac{3p+2}{12})\Gamma(\frac{3p+10}{12})(p+\frac{10}{3})$.
This case is the same as the well known eq. (6.53) of Rybicki $\&$
Lightman (1979). For $\nu_2\ll\nu$, we get
\begin{equation}
k_{\nu}=\frac{\sqrt{3}\pi^{3/2}}{9\sqrt{2}}\frac{(p+2)}{\sin{\theta}}N_{\gamma_e}
\gamma_2^{-(p+4)}\frac{q_e}{B}\left(\frac{\nu}{\nu_2}\right)^{-(p+4)/2}e^{-\nu/\nu_2},
\end{equation}
and if we take the isotropic distribution $<\sin{\theta}>=1/4$,
\begin{equation}
k_{\nu}=\frac{2\sqrt{6}\pi^{3/2}}{9}(p+2)N_{\gamma_e}\gamma_2^{-(p+4)}\frac{q_e}{B}
\left(\frac{\nu}{\nu_2}\right)^{-(p+4)/2}e^{-\nu/\nu_2},
\end{equation}
which is always neglected in astrophysics but it may be applied
for GRBs in which the self-absorption $\nu_a>\nu_2$ is possible
and it may explain the spectrum of GRBs combining inverse Compton
scattering.

For unification and convenience for comparison in the next section
we rewrite eqs. (A3), (A5) and (A7),
\begin{equation}
k_{\nu}=c_1\frac{q_e}{B}N_{\gamma_e}\gamma_1^{-(p+4)}\left(\frac{\nu}{\nu_1}\right)^{-5/3},
\nu\ll\nu_1
\end{equation}
\begin{equation}
k_{\nu}=c_2\frac{q_e}{B}N_{\gamma_e}\gamma_1^{-(p+4)}\left(\frac{\nu}{\nu_1}\right)^{-(p+4)/2},
\nu_1\ll\nu\ll\nu_2
\end{equation}
\begin{equation}
k_{\nu}=c_3\frac{q_e}{B}N_{\gamma_e}\gamma_2^{-(p+4)}\left(\frac{\nu}{\nu_2}\right)^{-(p+4)/2}e^{-\nu/\nu_2},
\nu_2\ll\nu,
\end{equation}
where $c_1=\frac{32\pi^2}{9\cdot
2^{\frac{1}{3}}\cdot\Gamma(\frac{1}{3})}\frac{p+2}{p+\frac{2}{3}}$,
$c_2=\frac{32\pi}{9}2^{\frac{p}{2}}g(p)$ and
$c_3=\frac{2\sqrt{6}\pi^{\frac{3}{2}}}{9}(p+2)$ which are of the
same order. For illustration, $c_1=14.78$, $c_2=17.80$ and
$c_3=13.64$ for $p=2.5$; $c_1=15.60$, $c_2=15.64$ and $c_3=12.12$
for $p=2.0$. So the above expression of $k_{\nu}$ still is valid
even when $\nu$ approaches $\nu_1$ or $\nu_2$. For simulations,
the discontinuity of $k_{\nu}$ can be removed if we constrain
$c_1=c_2=c_3/e$.

\subsection{Self-absorption frequency and the spectrums of fast and
slow cooling cases} Now we consider the absorption by electrons
with a broken power-law distribution. The break Lorentz factors
$\gamma_1<\gamma_2<\gamma_3$ with power law indices $p_1$ of
$(\gamma_1,\gamma_2)$ (hereafter we refer to this region as region
A) and $p_2$ of $(\gamma_2,\gamma_3)$ (region B). For slow cooling
$p_1=-p$ and $p_2=-(p+1)$ with $\gamma_1=\gamma_m$,
$\gamma_2=\gamma_c$, and $\gamma_3=\gamma_M$. For fast cooling
$p_1=-2$ and $p_2=-(p+1)$ with $\gamma_1=\gamma_c$,
$\gamma_2=\gamma_m$, and $\gamma_3=\gamma_M$. Both regions A and B
contribute to $k_{\nu}$. According to eqs. (A8), (A9) and (A10),
we find that $k_{\nu}$ is mainly determined by a nearer region
with respect to its frequency $\nu$, i.e., $k_{\nu}=k_{\nu}(A)$
for $\nu\leq\nu_1$ and $\nu_1<\nu<\nu_2$, while
$k_{\nu}=k_{\nu}(B)$ for $\nu_2<\nu\leq\nu_3$ and $\nu_3<\nu$.

The self-absorption frequency $\nu_a$ is determined by $k_{\nu_a}
L\approx 1$, where $L$ is the length of the radiation region in the comoving
frame. The spectrum below $\nu_a$ will be corrected since the optical
depth is larger than unity and $I_{\nu}\propto j_{\nu}/k_{\nu}$. From
the above analysis, we obtain the complete spectrum in all cases.

In the fast cooling phase, (1) for $\nu_a<\nu_c<\nu_m$ the
spectral indices of four segments are
$(2,\frac{1}{3},-\frac{1}{2},-\frac{p}{2})$; (2) for
$\nu_c<\nu_a<\nu_m$ the spectral indices are
$(2,\frac{5}{2},-\frac{1}{2},-\frac{p}{2})$; (3) for
$\nu_c<\nu_m<\nu_a(<\nu_M)$ the spectral indices are
$(2,\frac{5}{2},\frac{5}{2},-\frac{p}{2})$. Case (3) is previously
rarely discussed in astrophysics and maybe contributes to GRBs but
not to their afterglows. There is still case (4) $\nu_M<\nu_a$. We
do not discuss it since the flux of $\nu>\nu_M$ is negligible.

In the slow cooling phase, (1) for $\nu_a<\nu_m<\nu_c(<\nu_M)$ the
spectral indices are
$(2,\frac{1}{3},-\frac{p-1}{2},-\frac{p}{2})$; (2) for
$\nu_m<\nu_a<\nu_c(<\nu_M)$ the spectral indices are
$(2,\frac{5}{2},-\frac{p-1}{2},-\frac{p}{2})$; (3) for
$\nu_m<\nu_c<\nu_a(<\nu_M)$ the spectral indices are
$(2,\frac{5}{2},\frac{5}{2},-\frac{p}{2})$. For case (1), from eq.
(A4) we can find $\nu_a$ in the observer's frame is consistent
with eq. (22) of Wijers $\&$ Galama (1999), or we use eq. (A8) to
get the same $\nu_a$ as eq. (52) ($\nu <\nu_p$) of Panaitescu $\&$
Kumar (2000), if we take $N_{\gamma_e}\simeq (p-1) n_e
\gamma_m^{p-1}$ and $n_e$ is the electron density in the comoving
frame.
\end{appendix}

\newpage
\begin{figure}
  \begin{center}
  \centerline{ \hbox{\epsfig{figure=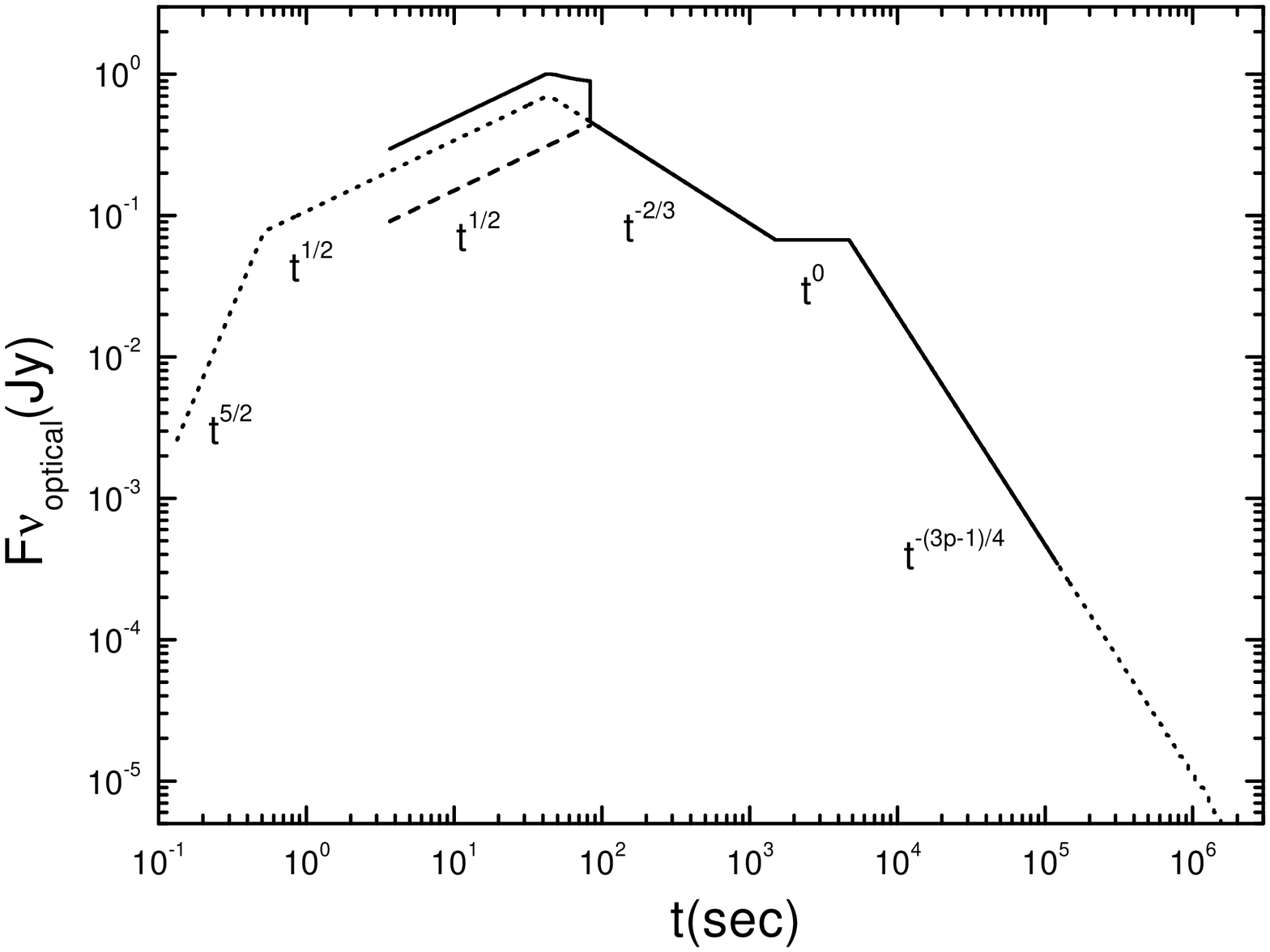,width=5.5in,height=3.8in,angle=0}}}
  \centerline{ \hbox{\epsfig{figure=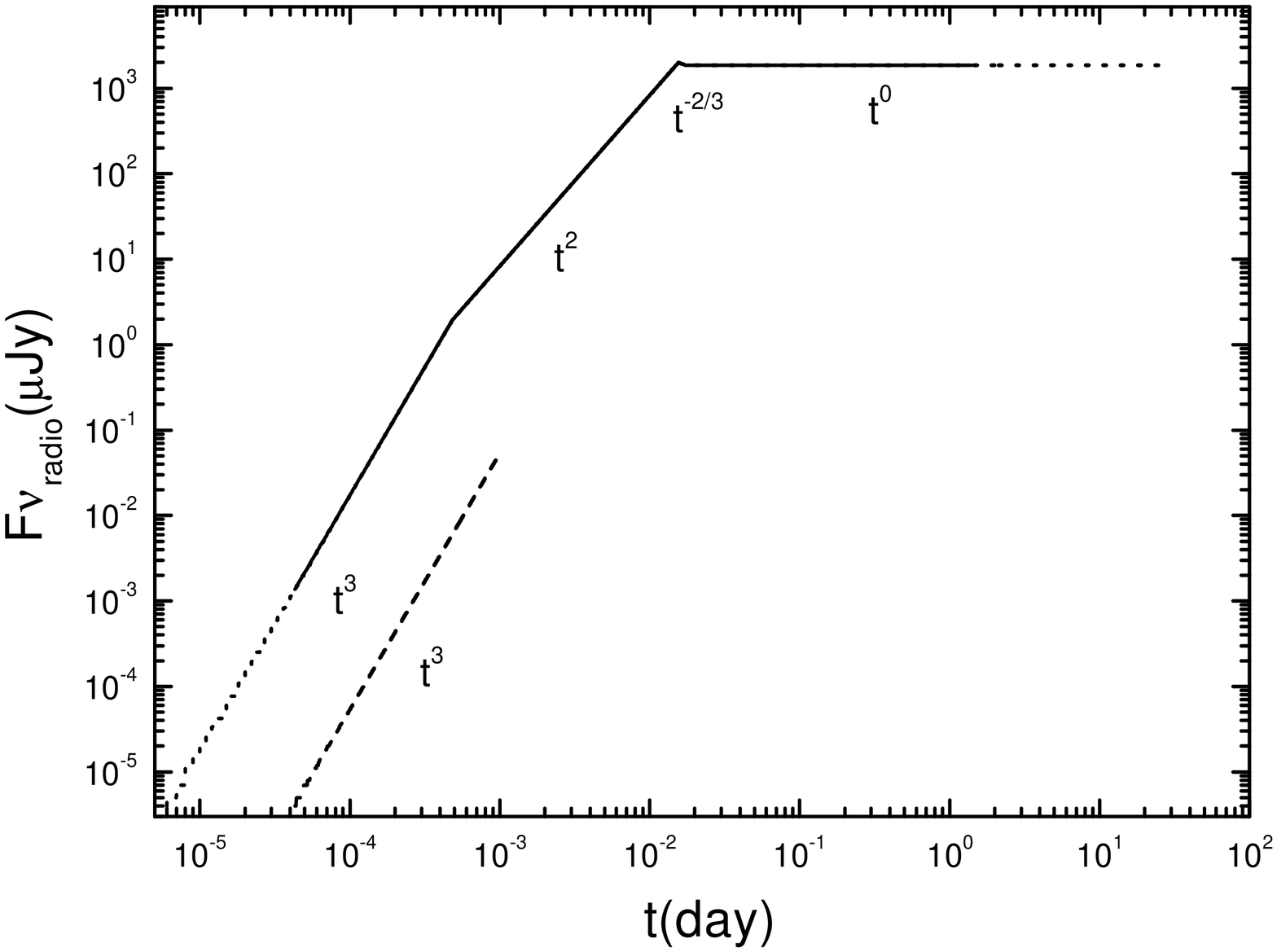,width=5.5in,height=3.8in,angle=0}}}
  \caption{Light curves of very early afterglows in a stellar wind.
The upper panel corresponds to optical flashes at
$\nu_{opt}=4\times 10^{14}$ Hz, and the bottom panel to radio
flares at $8.46$ GHz. The dashed, dotted and solid lines represent
the radiation from the relativistic reverse shock and forward
shock and their total emission. $E=10^{52}$ergs, $A=3\times
10^{34}$ cm$^{-1}$, $\eta=300$, $\Delta=5\times 10^{12}$cm,
$\xi_e=0.6$, $\xi_B=10^{-2}$, $D=10^{28}$cm and $p=2.5$ are
assumed.
 $t_{co}\approx856$sec is the time when $\nu_c=\nu_{opt}$.}
  \end{center}
  \end{figure}

\newpage
\begin{figure}
  \begin{center}
  \centerline{ \hbox{\epsfig{figure=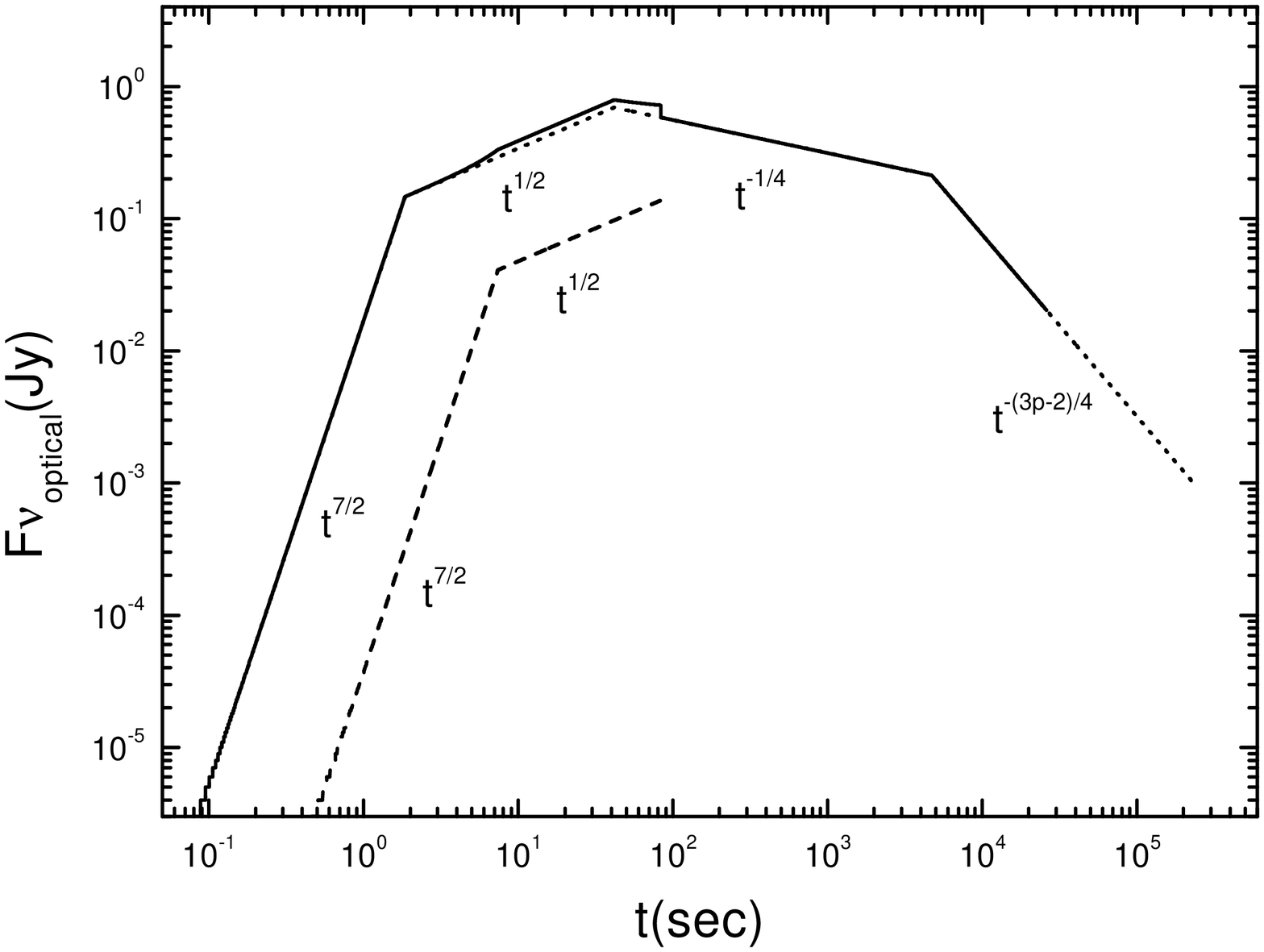,width=5.5in,height=3.8in,angle=0}}}
  \centerline{ \hbox{\epsfig{figure=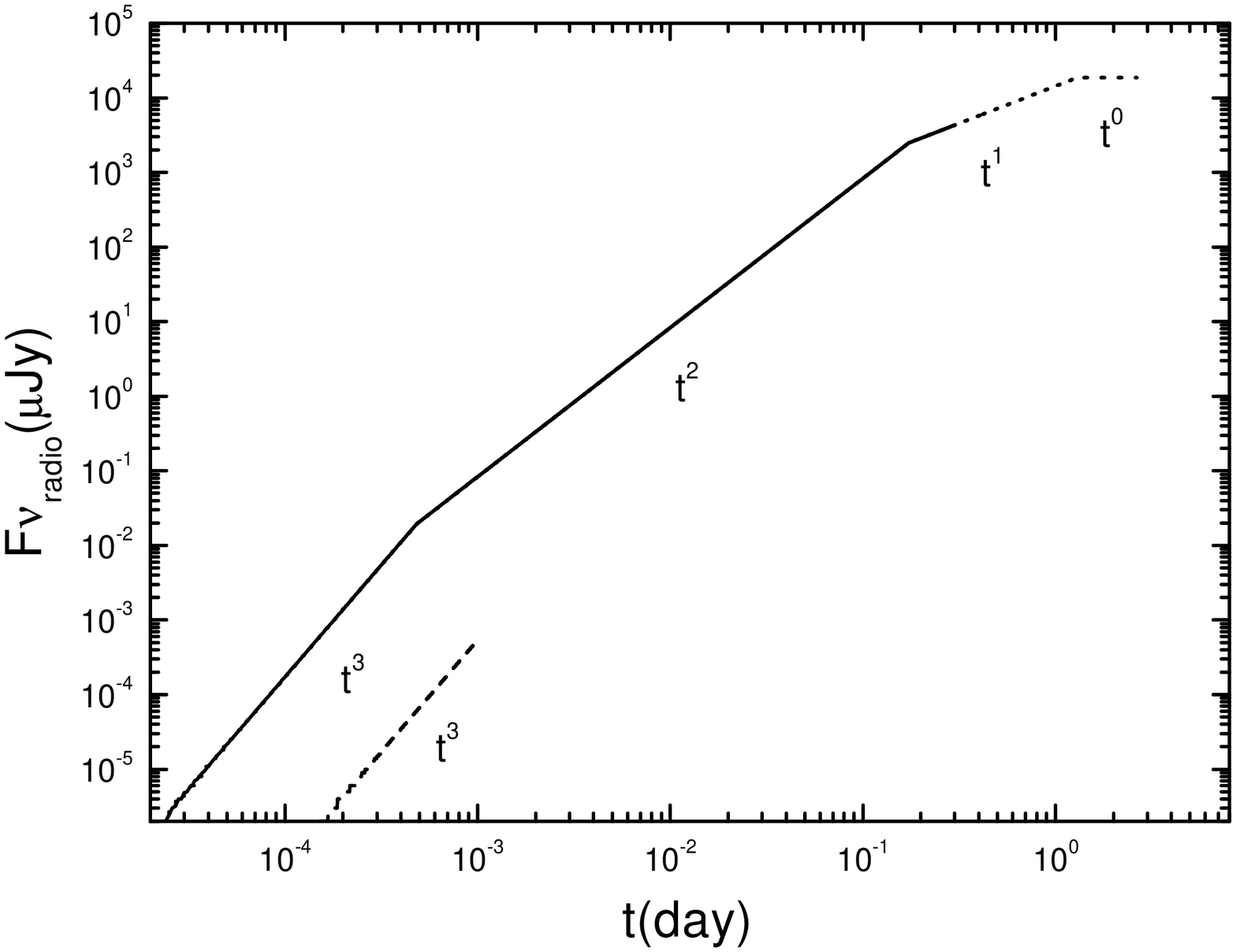,width=5.5in,height=3.8in,angle=0}}}
  \caption{Light curves of very early afterglows in a stellar wind.
Same as Fig.$1$ except for $E=10^{52}$ergs, $A=3\times 10^{35}$
cm$^{-1}$.}
  \end{center}
  \end{figure}

\newpage
\begin{figure}
  \begin{center}
  \centerline{ \hbox{\epsfig{figure=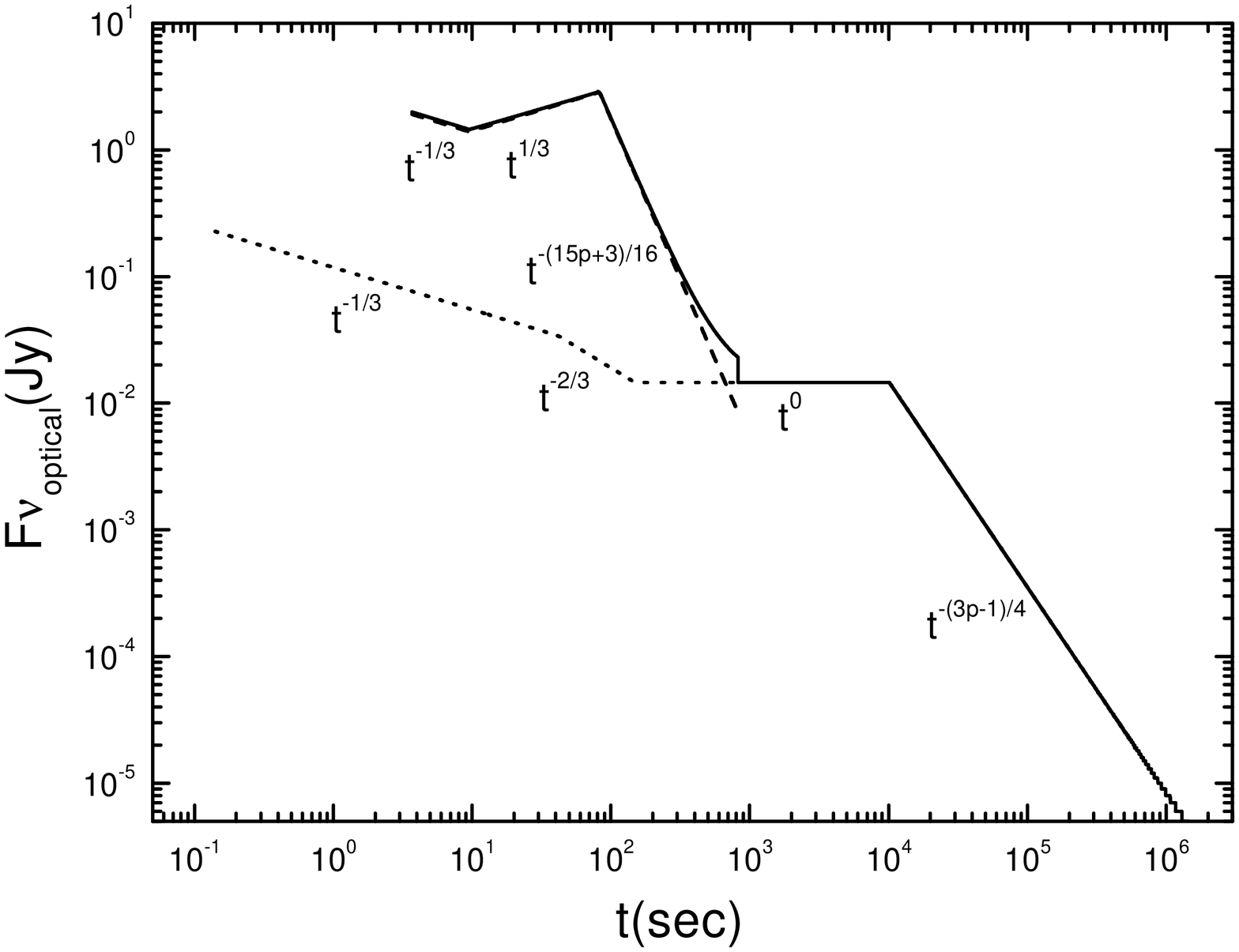,width=5.5in,height=3.8in,angle=0}}}
  \centerline{ \hbox{\epsfig{figure=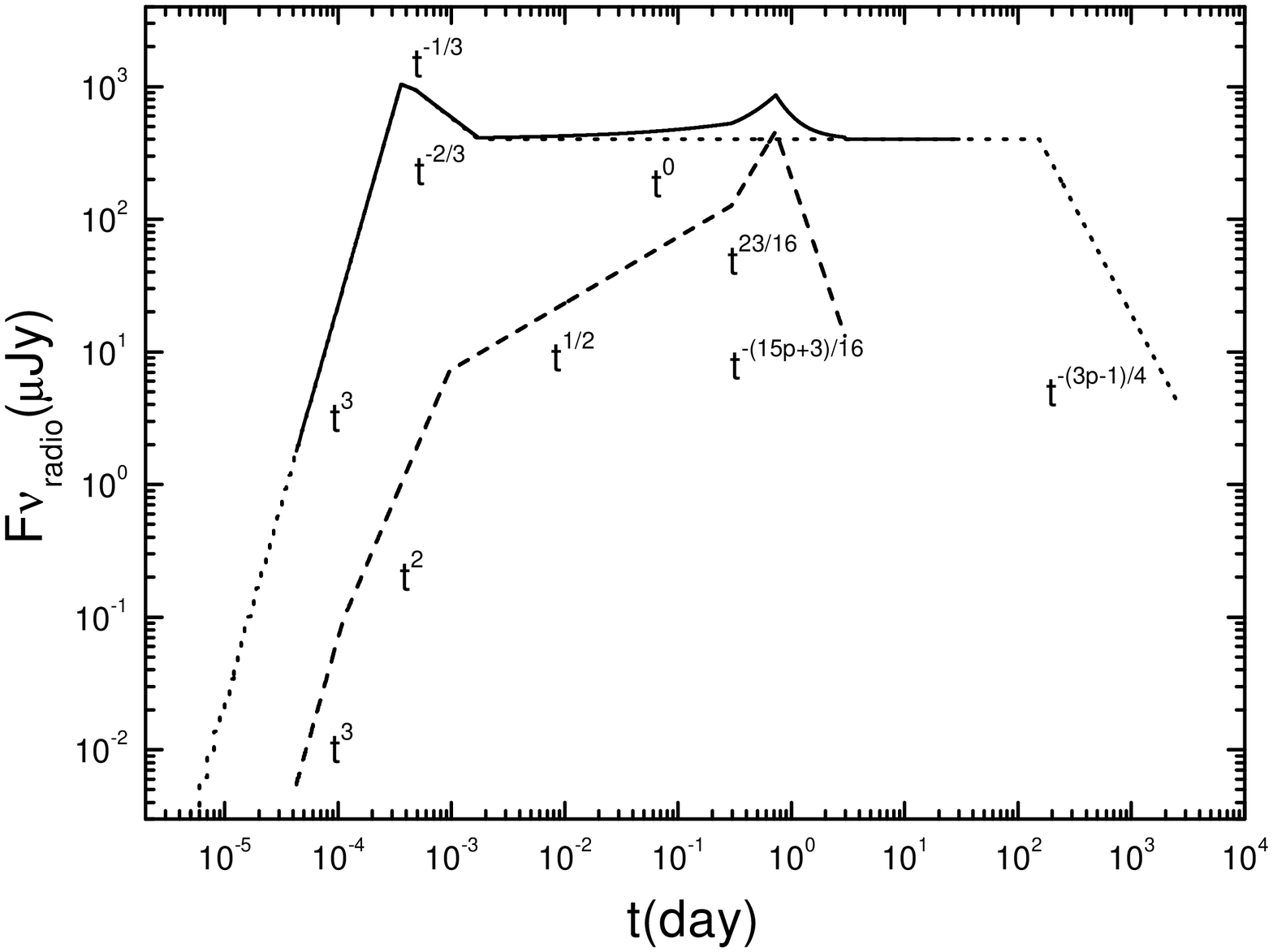,width=5.5in,height=3.8in,angle=0}}}
  \caption{Light curves of very early afterglows in a stellar wind.
Same as Fig.$1$ except for $E=10^{53}$ergs, $A=3\times 10^{33}$
cm$^{-1}$.}
  \end{center}
  \end{figure}

\newpage
\begin{figure}
  \begin{center}
  \centerline{ \hbox{ \epsfig{figure=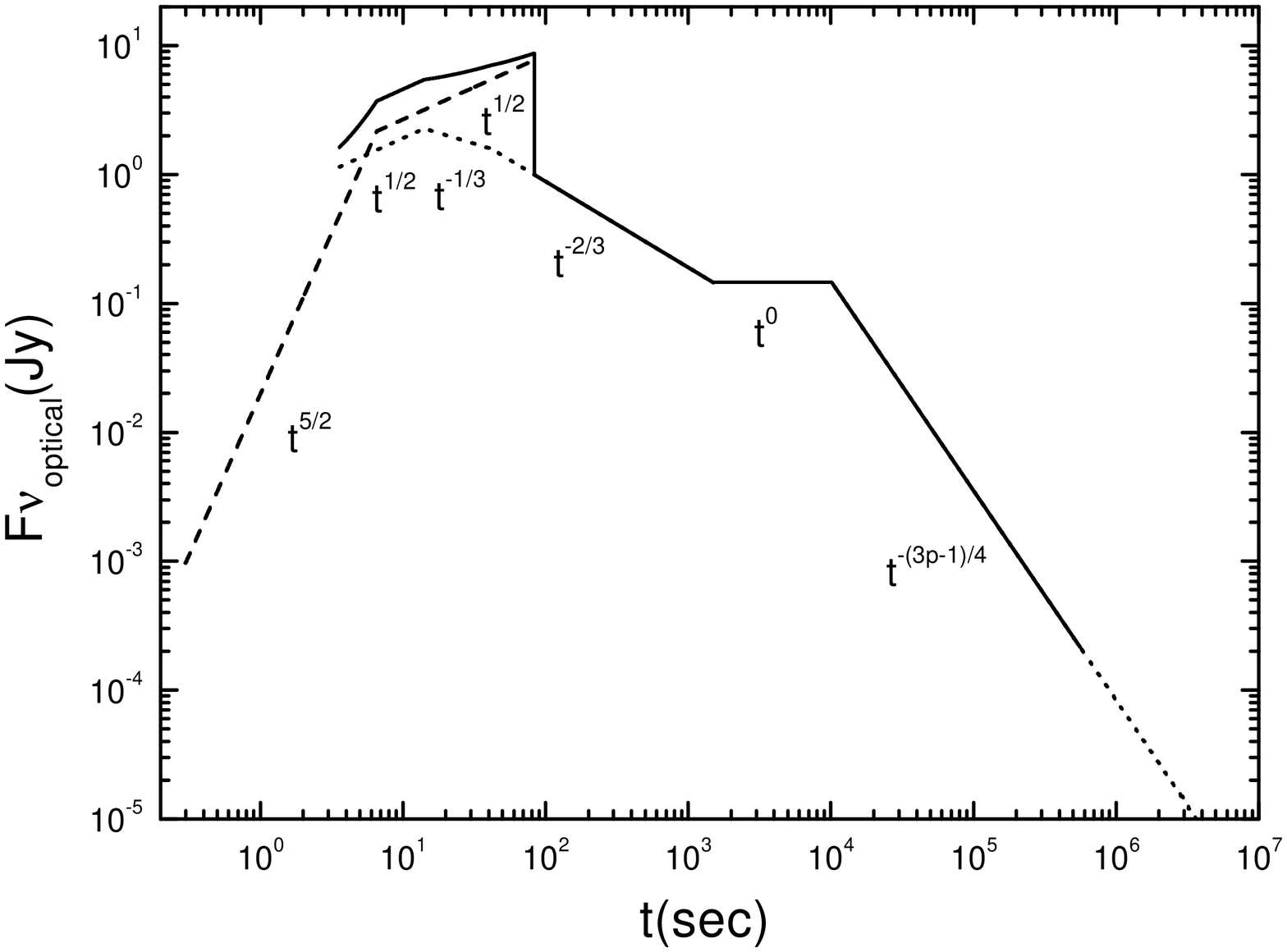,width=5.5in,height=3.8in,angle=0}}}
  \centerline{ \hbox{ \epsfig{figure=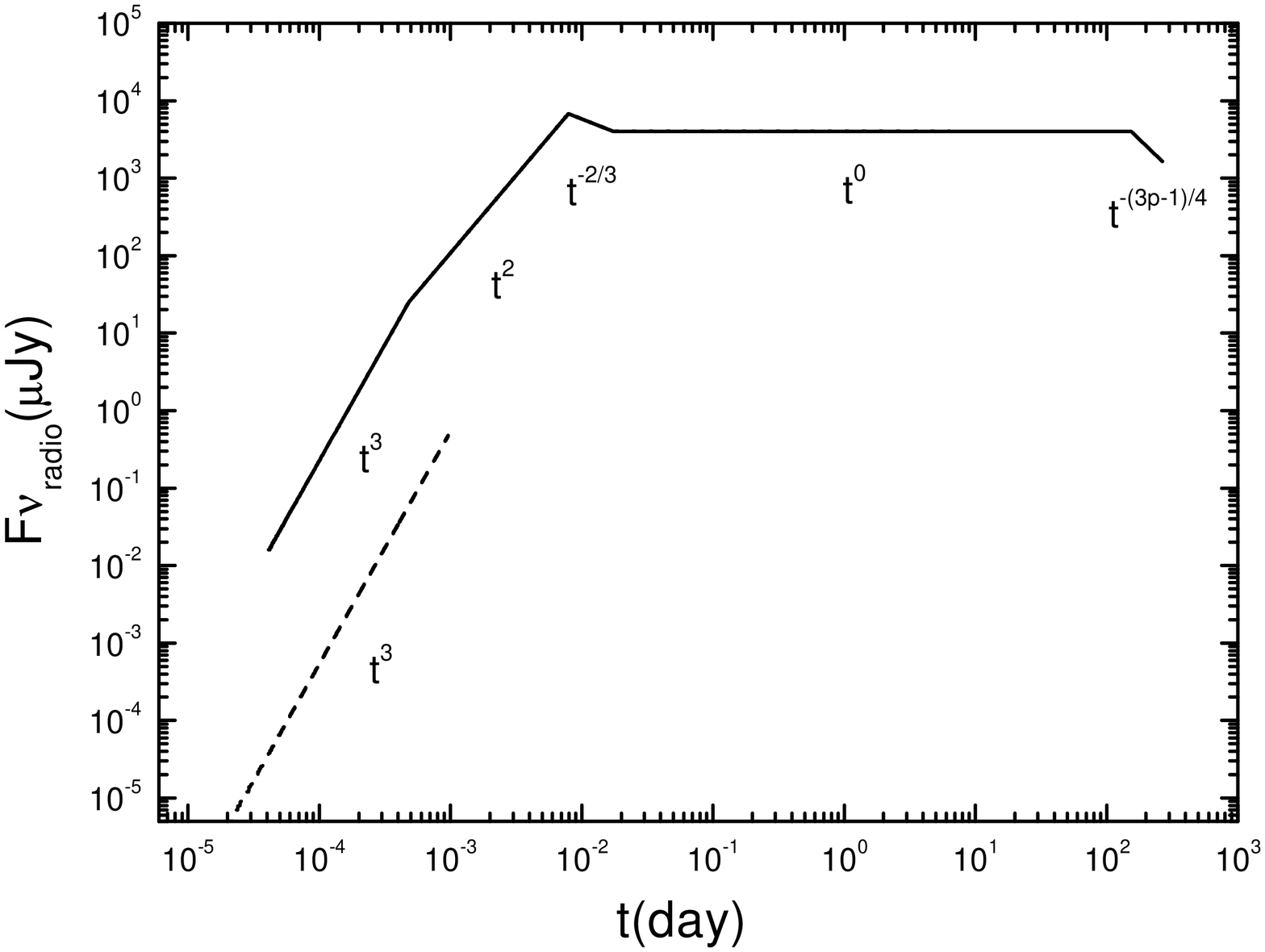,width=5.5in,height=3.8in,angle=0}}}
  \caption{Light curves of very early afterglows in a stellar wind.
Same as Fig.$1$ except for $E=10^{53}$ergs, $A=3\times 10^{34}$
cm$^{-1}$.}
  \end{center}
  \end{figure}

\newpage
\begin{figure}
  \begin{center}
  \centerline{ \hbox{ \epsfig{figure=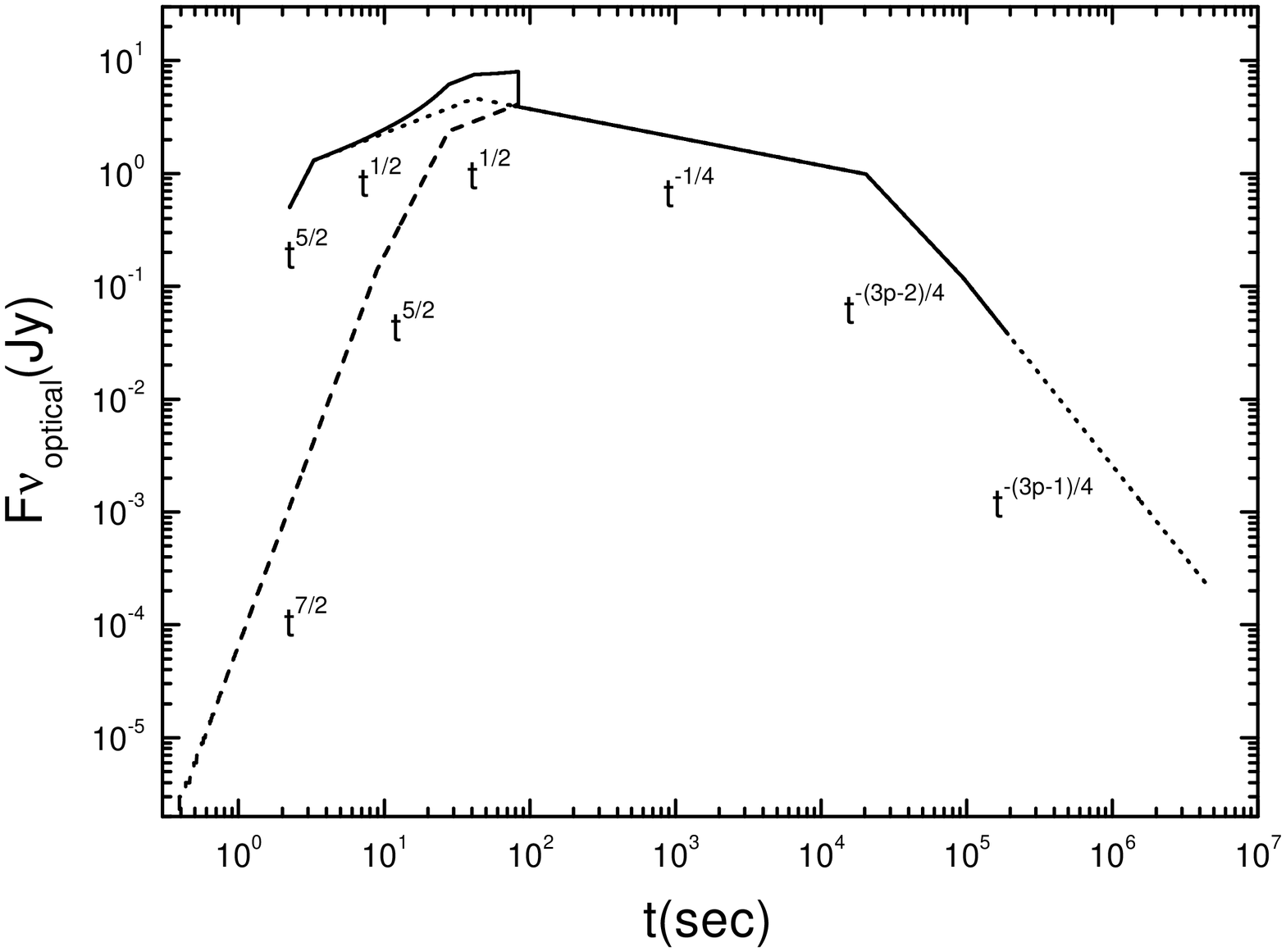,width=5.5in,height=3.8in,angle=0}}}
  \centerline{ \hbox{ \epsfig{figure=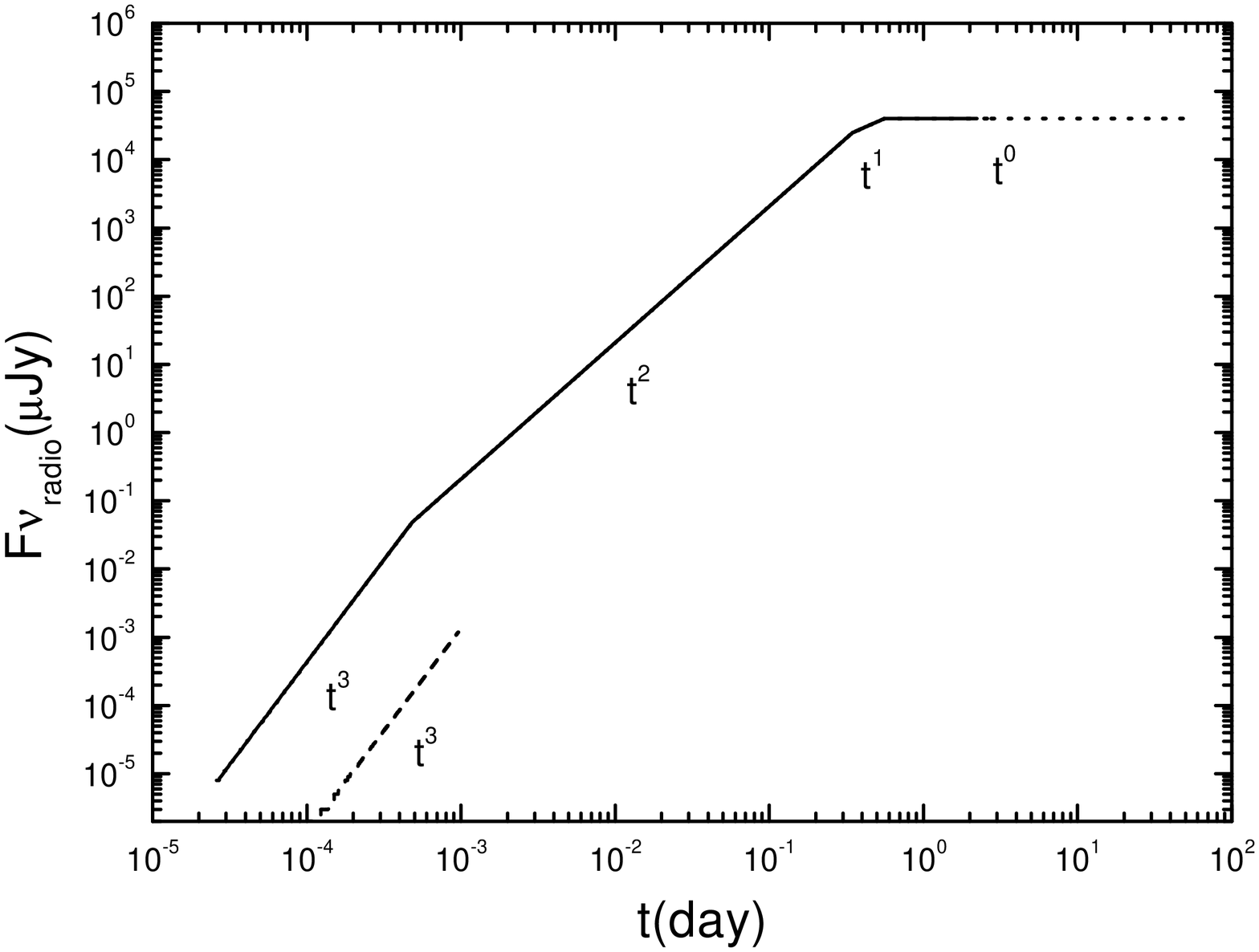,width=5.5in,height=3.8in,angle=0}}}
  \caption{Light curves of very early afterglows in a stellar wind.
Same as Fig.$1$ except for $E=10^{53}$ergs, $A=3\times 10^{35}$
cm$^{-1}$.}
  \end{center}
  \end{figure}

\newpage
\begin{figure}
  \begin{center}
  \centerline{ \hbox{ \epsfig{figure=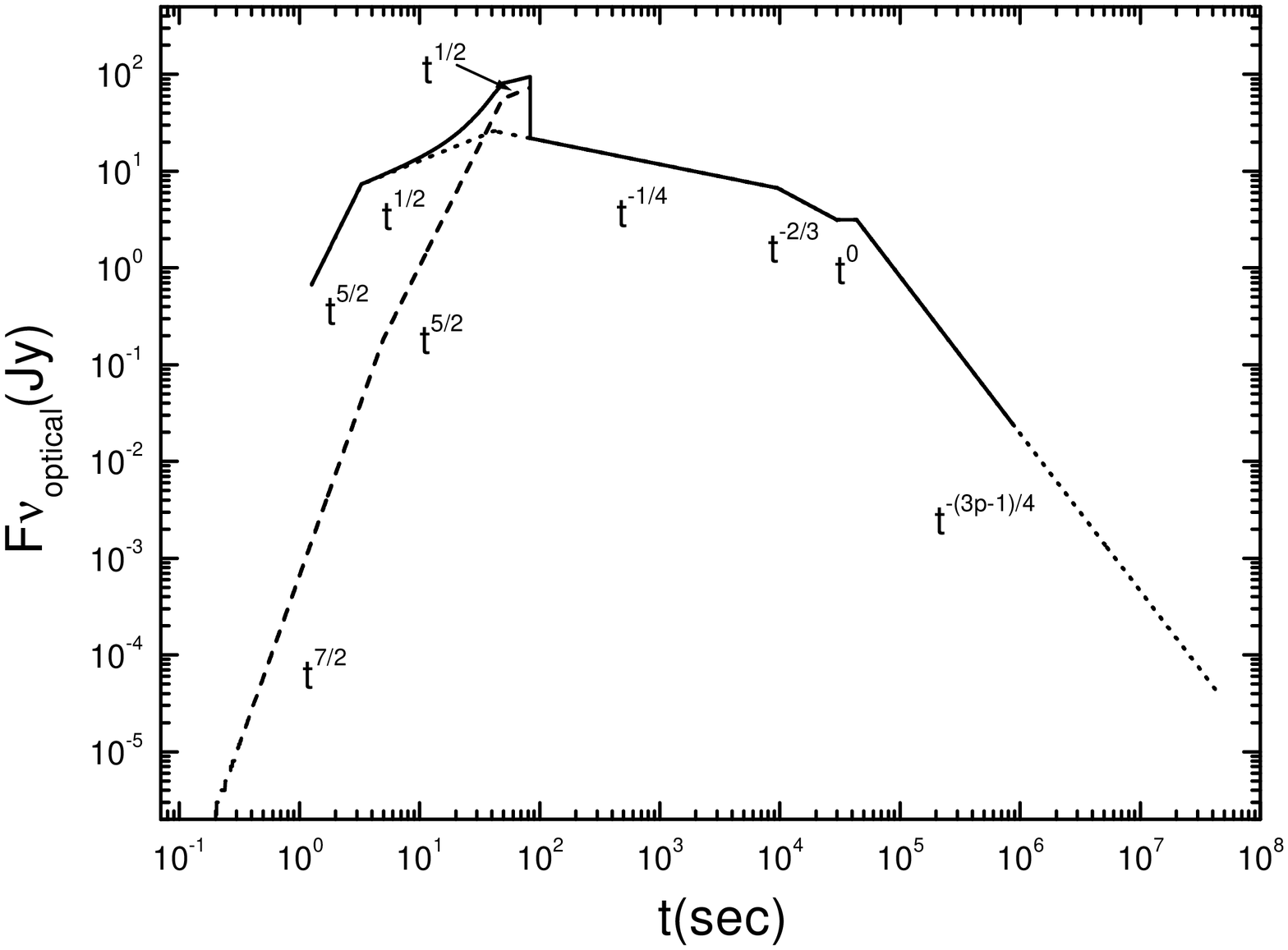,width=5.5in,height=3.8in,angle=0}}}
  \centerline{ \hbox{ \epsfig{figure=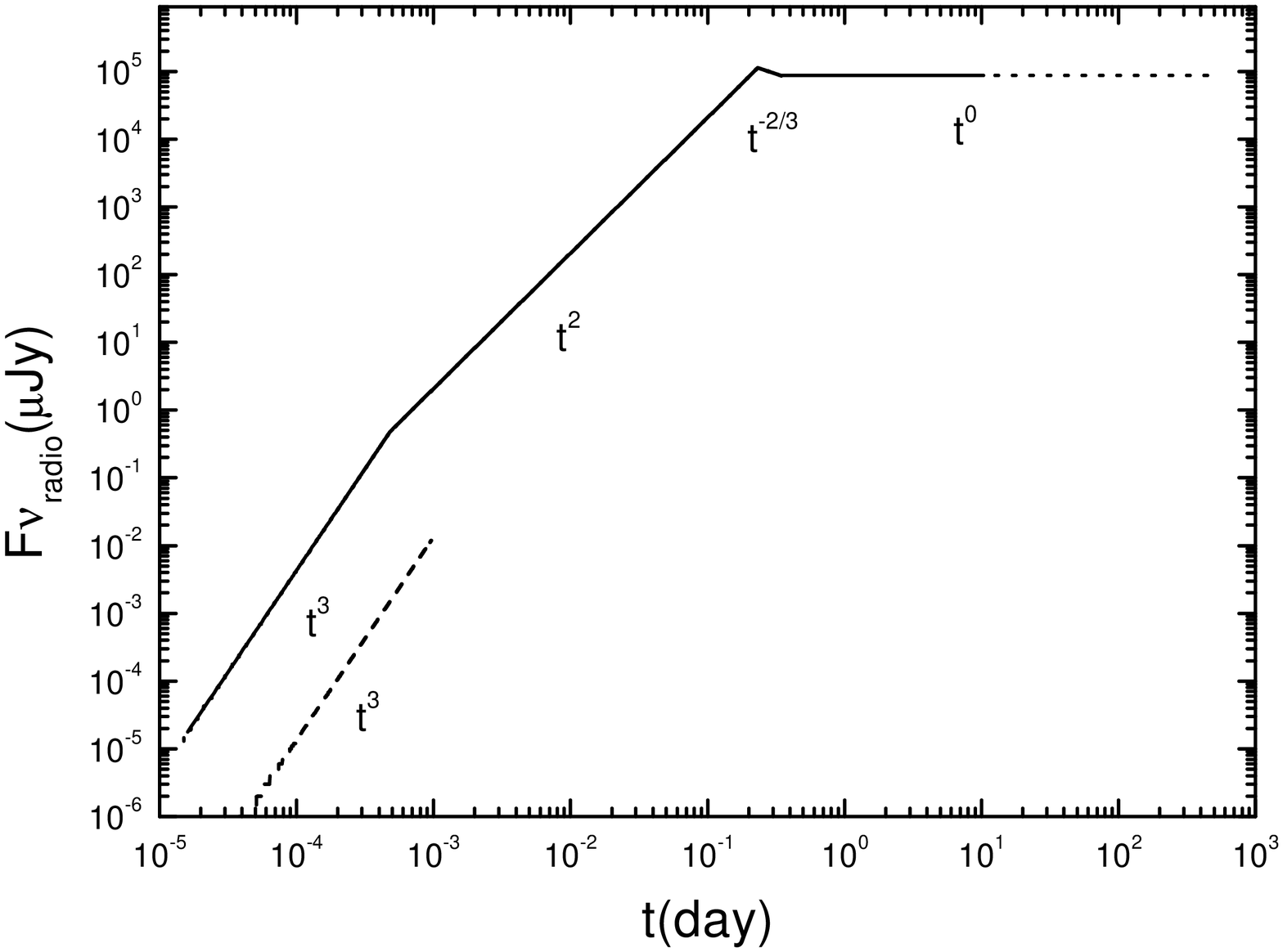,width=5.5in,height=3.8in,angle=0}}}
  \caption{Light curves of very early afterglows in a stellar wind.
Same as Fig.$1$ except for $E=10^{54}$ergs, $A=3\times 10^{35}$
cm$^{-1}$.}
  \end{center}
  \end{figure}

\end{document}